\documentclass[journal]{IEEEtran}

\usepackage{amsmath,amsfonts}
\usepackage{amssymb}
\usepackage{amsthm}
\usepackage{bm}
\usepackage{comment}
\usepackage{algorithmic}
\usepackage{array}
\usepackage[caption=false,font=normalsize,
labelfont=sf,textfont=sf]{subfig}
\usepackage{textcomp}
\usepackage{stfloats}
\usepackage{url}
\usepackage{verbatim}
\usepackage{graphicx}
\usepackage{balance}
\usepackage{booktabs}
\usepackage{tikz}
\usetikzlibrary{
    arrows.meta,
    positioning,
    calc,
    shapes.geometric,
    fit
}

\newtheorem{theorem}{Theorem}
\newtheorem{corollary}{Corollary}

\newtheorem{prop}{Proposition}

\title{On the Grouping-based RIS-aided Multi-user Networks: A Graph Theoretic Approach}

\author{Authors}
\author{Udit Raj and Priyadarshi Mukherjee, \textit{Senior Member, IEEE}
\thanks{The authors are with  the Department of Electrical Engineering and Computer Science, Indian Institute of Science Education and Research Bhopal, Bhopal, Madhya Pradesh 462066, India (e-mail: udit2026@iiserb.ac.in; priyadarshi@ieee.org).}\vspace{-12mm}}

\allowdisplaybreaks

\begin{document}

\maketitle

\begin{abstract}
In this letter, we propose a graph theoretic framework for grouping-based reconfigurable intelligent surface (RIS)-aided multi-user wireless networks. By considering both the ideal as well as the practical RIS-based phase shift models, the aspect of the RIS-user association is formulated as a graph matching problem. Thereafter, we derive conditions, which on satisfying, guarantee that the ideal RIS-based optimal RIS-user association remains uniquely optimal even in the practical RIS-based scenario. Finally, extensive Monte Carlo simulations validate the importance of the proposed framework.
\end{abstract}


\IEEEpeerreviewmaketitle

\section{Introduction}

The reconfigurable intelligent surface (RIS) is a promising technology for enhancing the performance of wireless networks \cite{8811733}. An RIS, consisting of reconfigurable passive elements, essentially reflects the incident signal in a desired direction by adjusting its parameters \cite{9140329}. Motivated by this, the interesting research direction of RIS-assisted device-to-device (D2D) communication \cite{tnse} has emerged in the recent years. Moreover, the adoption of high-frequency signals, such as millimeter waves, can facilitate high data rates. However, these type of signals suffer from drawbacks like high penetration and significant propagation losses. As a result, RIS-aided D2D networks appears to be the perfect solution in such scenarios, especially where the direct line of sight (LoS) link is weak.

Note that any given D2D pair communicates with each other for a finite period of time. Thus, employing an RIS for every D2D pair leads to wastage of resources. Accordingly, the authors in  \cite{partition3} suggest strategies, where the nearby elements of the RIS are grouped into smaller non-overlapping surfaces with each of them serving a dedicated D2D pair. In the context of a multi-user scenario, multiple D2D pairs intend to communicate by using a dedicated group of the RIS. Consequently, the aspect of D2D pair and RIS group association becomes crucial and it needs to be properly addressed. In this context, the authors in \cite{chan} investigate the aspect of jointly optimizing user association, beamforming, and RIS phase configuration in a multi-user RIS-aided network. The work in \cite{sautcom} proposes various group selection strategies for an RIS-assisted D2D communication set-up. Both these works consider an ideal RIS-based phase-shift model with unity reflection amplitude regardless
of the phase shift at each element. On the contrary, the authors in \cite{abeywickrama2020practical} propose a practical RIS-based phase shift model based on experimental results, where the reflection amplitude depends on the applied phase shift, and this has a significant impact on the system performance.

Motivated by this, we present a graph theoretic approach towards analyzing the grouping-based RIS-aided multi-user networks. The assignment of the RIS groups to the D2D pairs is formulated as a matching problem over a weighted bipartite graph. We derive conditions that determine whether the ideal RIS-based optimal matching remains uniquely optimal even in the practical RIS-based scenario. Finally, numerical results validate the proposed framework in the context of appropriate choice of system parameters. To the best of our knowledge, this is the first work that presents a complete graph theoretic framework for practical RIS-aided multi-user networks.

\section{System Model}

\begin{figure}[!t]
\centering

\resizebox{\linewidth}{!}{%
\begin{tikzpicture}[
    >=Latex,
    font=\small,
    every node/.style={font=\small},
    note/.style={draw,rounded corners,align=center,inner sep=3pt},
    apanchor/.style={minimum width=8mm,minimum height=10mm,inner sep=0pt},
    phoneanchor/.style={minimum width=7mm,minimum height=11mm,inner sep=0pt}
]

\node[font=\bfseries] at (1.2,7.4) {Transmitters Tx};

\node[apanchor] (T1) at (0.80,6.30) {};
\node[apanchor] (T2) at (1.55,5.15) {};
\node[apanchor] (T3) at (0.65,3.85) {};
\node[apanchor] (T4) at (1.40,2.65) {};

\draw[thick] (0.80,6.05) -- (0.80,6.45);
\fill (0.80,6.45) circle (1.2pt);
\draw[thick] (0.66,6.05) -- (0.80,5.90) -- (0.94,6.05);
\draw[thick] (0.85,6.45)
arc[start angle=0,end angle=55,radius=0.17];
\draw[thick] (0.75,6.45)
arc[start angle=180,end angle=125,radius=0.17];
\draw[thick] (0.88,6.45)
arc[start angle=0,end angle=60,radius=0.27];
\draw[thick] (0.72,6.45)
arc[start angle=180,end angle=120,radius=0.27];
\node[left=5mm of T1] {$T_1$};

\draw[thick] (1.55,4.90) -- (1.55,5.30);
\fill (1.55,5.30) circle (1.2pt);
\draw[thick] (1.41,4.90) -- (1.55,4.75) -- (1.69,4.90);
\draw[thick] (1.60,5.30)
arc[start angle=0,end angle=55,radius=0.17];
\draw[thick] (1.50,5.30)
arc[start angle=180,end angle=125,radius=0.17];
\draw[thick] (1.63,5.30)
arc[start angle=0,end angle=60,radius=0.27];
\draw[thick] (1.47,5.30)
arc[start angle=180,end angle=120,radius=0.27];
\node[left=5mm of T2] {$T_2$};

\draw[thick] (0.65,3.60) -- (0.65,4.00);
\fill (0.65,4.00) circle (1.2pt);
\draw[thick] (0.51,3.60) -- (0.65,3.45) -- (0.79,3.60);
\draw[thick] (0.70,4.00)
arc[start angle=0,end angle=55,radius=0.17];
\draw[thick] (0.60,4.00)
arc[start angle=180,end angle=125,radius=0.17];
\draw[thick] (0.73,4.00)
arc[start angle=0,end angle=60,radius=0.27];
\draw[thick] (0.57,4.00)
arc[start angle=180,end angle=120,radius=0.27];
\node[left=5mm of T3] {$T_3$};

\draw[thick] (1.40,2.40) -- (1.40,2.80);
\fill (1.40,2.80) circle (1.2pt);
\draw[thick] (1.26,2.40) -- (1.40,2.25) -- (1.54,2.40);
\draw[thick] (1.45,2.80)
arc[start angle=0,end angle=55,radius=0.17];
\draw[thick] (1.35,2.80)
arc[start angle=180,end angle=125,radius=0.17];
\draw[thick] (1.48,2.80)
arc[start angle=0,end angle=60,radius=0.27];
\draw[thick] (1.32,2.80)
arc[start angle=180,end angle=120,radius=0.27];
\node[left=5mm of T4] {$T_4$};

\node[font=\bfseries] at (6.6,7.4) {RIS};

\coordinate (SW) at (4.0,2.0);
\coordinate (NE) at (9.2,7.2);

\fill[blue!18] (4.0,4.6) rectangle (6.6,7.2);
\fill[gray!10] (6.6,4.6) rectangle (9.2,7.2);
\fill[gray!10] (4.0,2.0) rectangle (6.6,4.6);
\fill[gray!10] (6.6,2.0) rectangle (9.2,4.6);

\draw[thick] (SW) rectangle (NE);
\draw[thick] (6.6,2.0) -- (6.6,7.2);
\draw[thick] (4.0,4.6) -- (9.2,4.6);

\node[blue!80!black,font=\bfseries] at (5.3,6.5) {$G_1$};
\node at (7.9,5.9) {$G_2$};
\node at (5.3,3.3) {$G_3$};
\node at (7.9,3.3) {$G_4$};

\foreach \x in
{4.2,4.55,4.9,5.25,5.6,5.95,6.8,7.15,7.5,7.85,8.2,8.55}
{
    \foreach \y in
    {2.2,2.55,2.9,3.25,3.6,3.95,4.8,5.15,5.5,5.85,6.2,6.55}
    {
        \draw[gray!40] (\x,\y) rectangle ++(0.22,0.22);
    }
}

\draw[very thick,red] (4.55,5.15) rectangle ++(0.22,0.22);
\draw[very thick,red] (5.05,5.75) rectangle ++(0.22,0.22);
\draw[very thick,red] (5.65,6.15) rectangle ++(0.22,0.22);

\node[red!80!black,left] at (4.52,5.26) {$n_i$};
\node[red!80!black,left] at (5.02,5.86) {$n_j$};
\node[red!80!black,right] at (5.90,6.26) {$n_k$};

\node[font=\bfseries] at (12.3,7.4) {Receivers Rx};

\node[phoneanchor] (R1) at (12.70,6.10) {};
\node[phoneanchor] (R2) at (11.75,5.00) {};
\node[phoneanchor] (R3) at (12.85,3.75) {};
\node[phoneanchor] (R4) at (11.95,2.55) {};

\foreach \u/\lab in {R1/R_1,R2/R_2,R3/R_3,R4/R_4}
{
    \draw[thick,rounded corners=1pt]
    ($(\u.center)+(-0.16,-0.30)$)
    rectangle
    ($(\u.center)+(0.16,0.30)$);

    \draw
    ($(\u.center)+(-0.11,-0.20)$)
    rectangle
    ($(\u.center)+(0.11,0.19)$);

    \draw[thick]
    ($(\u.center)+(-0.045,0.24)$)
    --
    ($(\u.center)+(0.045,0.24)$);

    \fill ($(\u.center)+(0,-0.25)$) circle (0.8pt);

    \node[right=4mm of \u] {$\lab$};
}

\draw[->,thick,blue!80] (T1.east) -- (4.55,5.26);
\draw[->,thick,blue!80] (T1.east) -- (5.16,5.86);
\draw[->,thick,blue!80] (T1.east) -- (5.76,6.26);

\draw[->,thick,blue!80] (4.77,5.26) -- (R1.west);
\draw[->,thick,blue!80] (5.27,5.86) -- (R1.west);
\draw[->,thick,blue!80] (5.87,6.26) -- (R1.west);

\node[blue!80!black] at (2.55,6.65)
{$g_{1n}$};

\node[blue!80!black] at (10.95,6.55)
{$h_{n1}$};

\draw[->,gray!50,dashed] (T2.east) -- (7.7,5.7);
\draw[->,gray!50,dashed] (7.9,5.7) -- (R2.west);

\draw[->,gray!50,dashed] (T3.east) -- (5.4,3.3);
\draw[->,gray!50,dashed] (5.6,3.3) -- (R3.west);

\draw[->,gray!50,dashed] (T4.east) -- (8.0,3.0);
\draw[->,gray!50,dashed] (8.2,3.0) -- (R4.west);

\end{tikzpicture}%
}
\vspace{-6mm}
\caption{Proposed system architecture with $4$ Tx-Rx pairs, i.e., here $K=4$.}
\label{fig:RIS_edge}
\vspace{-6mm}
\end{figure}
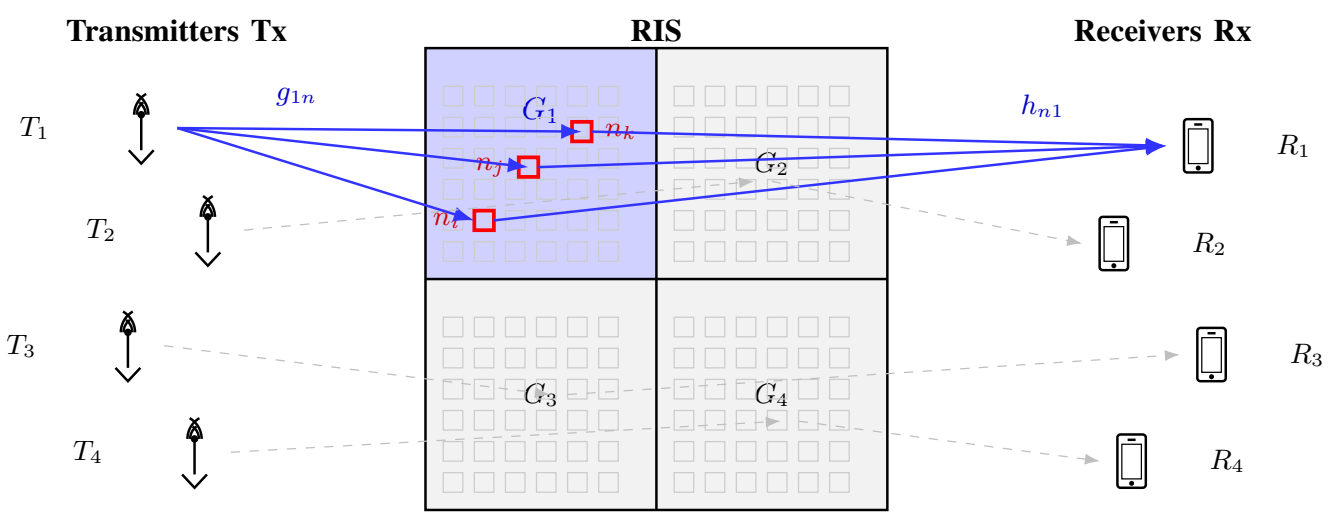

As in Fig. \ref{fig:RIS_edge}, we consider an RIS-aided  multi-user set-up. An individual transmitter (Tx) communicates with its dedicated receiver (Rx), each equipped with single antenna, and there are $K$ such pairs. We assume that no Tx-Rx pair is able to communicate directly and an RIS, consisting of $N_T$ reflecting elements, is used to complete the communication process.

\subsection{RIS Characterization}
The reflection coefficient of the $n$-th RIS element is expressed as $v_n=\beta_n e^{j\theta_n}$, where $\beta_n$ and $\theta_n$ denote the amplitude and phase shift, respectively. Ideally, we assume $\beta_n=1,\theta_n \in [0,2\pi)$ $\forall$ $n=1,\dots,N_T$. However in reality, $\beta_n$ and $\theta_n$ are coupled \cite{abeywickrama2020practical} and the reflection coefficient $v_n$ is modeled as $v_n=\beta_n(\theta_{n})e^{j\theta_{n}}$, where
\vspace{-2.4mm}
\begin{equation}  \label{btrel}
\beta_n(\theta_{n})= (1-\beta_{\min})
\left(\frac{\sin(\theta_{n}-\phi)+1}{2}
\right)^{\alpha}+\beta_{\min}.
\end{equation}
Here $\beta_{\min}$, $\alpha$, and $\phi$ are RIS circuit-dependent. The adjacent reflecting elements are assumed to be uncorrelated, i.e., a half-wavelength of inter-element spacing. To facilitate multiple Tx-Rx pairs by a single RIS, we employ a frequency division multiplexing-aided grouping-based strategy with each group catering to a different Tx-Rx pair. The RIS is partitioned into $K$ non-overlapping groups $R_q$ $\forall$ $1 \leq q \leq K$ with $N$ elements in each such that $N \times K=N_T$ \footnote{As we focus on the Tx-Rx pair and RIS group association, we are not assuming that all the elements belonging to the same group introduce identical phase shift. Without any loss of generality, various RIS grouping strategies \cite{partition3} and their impact on the proposed framework can also be investigated.} Based on the channel knowledge, the $q$-th RIS group provides the desired phase shift to the signal from the $k$-th Tx by varying the diagonal matrix
\vspace{-2.4mm}
\begin{equation}  \label{phidef}
    \Phi_{kq} ={\rm diag} \left( \beta_{k1}(\theta_{k1})e^{j\theta_{k1}},\cdots,\beta_{kN}(\theta_{kN})e^{j\theta_{kN}} \right).
\end{equation}

\subsection{Channel Model}
We assume that the wireless links experience both large-scale path loss effects and small scale fading. The received power at the $n$-th RIS element $(n=1,\dots,N_T)$ from the $k$-th Tx $(k=1,\dots,K)$ is attenuated by $L_{kn}^{\rm TR}=L_0 \left( d_{kn}^{\rm TR} \right)^{-\alpha_{\rm TR}}$, where $L_0$ is the path loss at a reference distance, $d_{kn}^{\rm TR}$ is the distance between the $k$-th Tx and the $n$-th RIS element, and $\alpha_{\rm TR}$ is the path loss exponent \cite{tnse}. The received signal, at the $n$-th RIS element from the $k$-th Tx, is
\vspace{-2.4mm}
\begin{equation}  \label{chmod}
    y_{kn}=\sqrt{P_{t,k}L_{kn}^{\rm TR}}g_{kn}s_k + w_{n},
\end{equation}
where $P_{t,k}$ is the transmission power of the $k$-th Tx, $g_{kn}$ is the flat fading complex channel coefficient, $s_k$ is the transmitted signal from the $k$-th Tx with $\mathbb{E}[|s_k|^2]=1$, and $w_{n}$ is the additive white Gaussian noise (AWGN) with zero mean and variance $\sigma_n^2$. We assume that $|g_{kn}|=\alpha_{kn}$ is a Rayleigh random variable \cite{wcl3} with $\mathbb{E}[\alpha_{kn}^2]=1$ $(k=1,\dots,K \quad n=1,\dots,N_T)$. Similarly, the received power at the $k$-th Rx is attenuated by $\displaystyle L_{nk}^{\rm RR}=L_0 \left( d_{nk}^{\rm RR} \right)^{-\alpha_{\rm RR}}$, with $d_{nk}^{\rm RR}$ being the distance between the $n$-th RIS element and the $k$-th Rx and $\alpha_{\rm RR}$ is the path loss exponent. Lastly, here also we model the channel between the RIS and the Rx as flat fading with complex coefficient $h_{nk}$ $(k=1,\dots,K \quad n=1,\dots,N_T)$, where $|h_{nk}|=\zeta_{nk}$ is Rayleigh distributed with $\mathbb{E}[\zeta_{nk}^2]=1$.

\section{Problem Formulation}

Here, we discuss the analytical characterization of our considered set-up and formulate the user-RIS association problem.

\subsection{Considered System Design}

Based on \eqref{chmod}, the resulting signal vector from the $k$-th Tx to the $q$-th group $(k,q=1,\dots,K)$ of the RIS is
\vspace{-2.4mm}
\begin{equation}  \label{yvec}
\boldsymbol{y_{kq}}=\sqrt{P_{t,k}L_{kq}^{\rm TR}}\boldsymbol{g_{kq}}s_k + \boldsymbol{w_{q}},
\end{equation}
where $\boldsymbol{y_{kq}}=[y_{k1},\cdots,y_{kN}]^T,$ $\boldsymbol{g_{kq}}=[g_{k1},\cdots,g_{kN}]^T,$ and $\boldsymbol{w_{q}}=[w_1,\cdots,w_N]^T$. As $\boldsymbol{y_{kq}}$ is reflected to the $k$-th Rx, the received signal at the $k$-th Rx is
\vspace{-2.4mm}
\begin{equation}  \label{yrx}
    y_{qk}=\sqrt{L_{qk}^{\rm RR}}\boldsymbol{h_{qk}}^T \Phi_{kp} \boldsymbol{y_{kp}} + z_k,
\end{equation}
where $\boldsymbol{h_{qk}}=[h_{1k},\cdots,h_{Nk}]^T$ and $z_k$ is the AWGN at the $k$-th Rx with zero mean and variance $\sigma_k^2$. Hence, by using \eqref{phidef} and \eqref{yvec} in \eqref{yrx}, we obtain
\vspace{-2.4mm}
\begin{align}
    y_{qk} & = \delta \sum_{i=1}^N \zeta_{ik}\alpha_{ki}\beta_{ki}\left( \theta_{ki} \right)e^{j (\theta_{ki}+\measuredangle h_{i,k}+\measuredangle g_{ki})}s_k + z_k,
\end{align}
where we define $\delta=\sqrt{P_{t,k}L_{kq}^{\rm TR}L_{qk}^{\rm RR}}$. Moreover, the noise at the RIS, i.e., $\boldsymbol{w_{q}}$ is ignored as it is negligible compared to $z_k$ after the reflection and going through the channel between the RIS and the $k$-th Rx. Therefore, with an optimal phase shift, i.e., $\theta_{ki}+\measuredangle h_{i,k}+\measuredangle g_{ki}=0$ at all the elements of the $q$-th group of the RIS, we obtain
\vspace{-2.4mm}
\begin{equation}  \label{yopt}
    y_{qk}^{\rm opt}=\delta \sum_{i=1}^N \zeta_{ik}\alpha_{ki}\beta_{ki}\left( \theta_{ki} \right)s_k + z_k.
\end{equation}
We observe from \eqref{btrel} that, since $\beta_{ki}\left( \theta_{ki} \right) \leq 1$ $\forall$ $i$, $y_{qk}^{\rm opt}$ is a function of $\theta_{k,1}\dots,\theta_{k,N}$. Accordingly, when the $q$-th RIS group is selected to connect the $k$-th Tx-Rx pair, the resulting signal-to-noise ratio (SNR) is
\vspace{-2.4mm}
\begin{equation}  \label{snrdef}
\gamma_{qk}=\frac{P_{t,k}L_{kq}^{\rm TR}L_{qk}^{\rm RR}}{\sigma_k^2} \left( \sum_{i=1}^N \zeta_{ik}\alpha_{ki}\beta_{ki}\left( \theta_{ki} \right) \right)^2.
\end{equation}

\subsection{Problem Formulation}

Our objective is to propose a framework that investigates the impact of the coupled amplitude and phase shift constraint at the RIS on the user-RIS association. To this end, we model the problem as a graph matching problem with $T_1,\dots,T_K$ denoting the $K$ Tx, $G_1,\dots,G_K$ the $K$ RIS groups, and $R_1,\dots,R_K$ the $K$ Rx. Here, the Tx-Rx pairing is fixed, i.e., $T_k$ communicates only with its corresponding $R_k$ $(k=1,\dots,K)$. Hence, for each pair $P_k=(T_k,R_k)$, the assignment problem only determines which $G_q$ $(q=1,\dots,K)$ serves $P_k$ and  cross-pair links such as $T_i\rightarrow G_q\rightarrow R_k$  $(i\neq k)$ are not considered.

Traditionally, a graph $\mathcal{G}=(V,E)$ is defined by a vertex set $V$ and an edge set $E$, where each edge $e\in E$ connects a pair of vertices and represents an interaction between that pair of vertices \cite{bipartite}. Here, $V$ consists of the $K$ RIS groups and the $K$ Tx-Rx pairs, i.e., $V=\{G_1,\ldots,G_K,P_1,\ldots,P_K\}$, which is partitioned into the RIS group set $\mathcal{V}_G=\{G_1,\ldots,G_K\}$ and the Tx-Rx pairs set $\mathcal{V}_P=\{P_1,\ldots,P_K\}.$ The edge set is defined as
$E=\{e_{qk}:q,k\in[K]\},$ where each edge $e_{qk}$ connects RIS group $G_q$ to the Tx-Rx pair $P_k$. Corresponding to $e_{qk}$, we define the associated edge weight $w_{qk}$ as the achievable spectral efficiency obtained by assigning $G_q$ to $P_k$, i.e., $w_{qk}=\log_2 \left( 1+ \gamma_{pk} \right)$. Specifically, we take into account two different scenarios. First, we have the ideal case of $\beta_n=1$ $\forall$ $n$ and second, we consider the practical case, where $\beta_n$ and $\theta_n$ are coupled as described in \eqref{btrel}. Accordingly, we define $\gamma_0=\frac{P_{t,k}L_{kq}^{\rm TR}L_{qk}^{\rm RR}}{\sigma_k^2}$ and by using \eqref{snrdef}, we obtain
\vspace{-2.4mm}
\begin{align}  \label{eweight}
   w_{qk}^{\rm ideal}&= \log_2 \left( 1 + \gamma_0 \left( \sum_{i=1}^N \zeta_{ik}\alpha_{ki} \right)^2 \right) \nonumber \\
   w_{qk}^{\rm prac}&= \log_2 \left( 1 + \gamma_0 \left( \sum_{i=1}^N \zeta_{ik}\alpha_{ki}\beta_{ki}\left( \theta_{ki} \right) \right)^2 \right).
\end{align}
Here, $w_{qk}^{\rm ideal}$ and $w_{qk}^{\rm prac}$ denote the edge weight  corresponding to the ideal and practical scenario, respectively. In this graph $\mathcal{G}$, edges exist only between $\mathcal{V}_G$ and $\mathcal{V}_P$, i.e., $\mathcal{G}$ is bipartite \cite{bipartite}. Moreover, since every RIS group is connected to every Tx-Rx pair, $\mathcal{G}$ is a complete bipartite graph containing $K^2$ edges. Note that, the graphs corresponding to the ideal and practical RIS-based models have the same topology but differ in their edge weights. Our objective is to associate exactly one group of the RIS with each of the $K$ Tx-Rx pairs such that the entire system performance, in terms of the sum spectral efficiency (SSE), is enhanced\footnote{Note that, here we consider a scenario with equal number of Tx-Rx pairs and groups in the RIS. As an immediate extension, we intend to investigate a generalized scenario with unequal number of Tx-Rx pairs and RIS groups.}.

\subsection{Illustrative Example}

We consider a scenario with $K=4$ and $V=\{G_1,\ldots,$ $G_4,P_1,\ldots,P_4\}$, i.e., $\mathcal{V}_G=\{G_1,\ldots,G_4\}$ and $\mathcal{V}_P=\{P_1,\ldots,P_4\}$.
The weighted adjacency matrix \cite{bipartite}, for the ideal RIS-based scenario, is $A^{\rm I}
=
\begin{bmatrix}
\mathbf{0}_{K\times K} & W^{\rm I}\\
\left(W^{\rm I}\right)^{T} & \mathbf{0}_{K\times K}
\end{bmatrix}_{2K\times 2K},$
where, $W^{\rm I}=\left[w_{qk}^{\rm ideal}\right]_{q,k=1}^{K}$ is the ideal edge-weight matrix and $\mathbf{0}_{K\times K}$ is the zero matrix of order $K$. The zero diagonal blocks and the non-zero off-diagonal blocks of $A^{\rm I}$ justify its bipartite nature, where the upper-right block is the weight matrix $W^{\rm I}=[w_{qk}^{\rm ideal}]$ and the lower-left block is $(W^{\rm I})^{T}$. The practical weighted adjacency matrix is analogously obtained by replacing $W^{\rm I}$ with $W^{\rm P}$. 

A \emph{matching} $\mathcal{M}\subseteq E$ is a set of edges such that no two edges in $\mathcal{M}$ share a common vertex, i.e., each RIS group and each Tx-Rx pair can be incident to at most one edge belonging to the matching. Moreover, it is called a \emph{perfect matching} if every vertex of $\mathcal{G}$ is incident to exactly one edge in the matching. Since $\mathcal{G}$ contains $K$ RIS-group vertices and $K$ Tx-Rx pair vertices, every perfect matching will contain exactly $K$ edges and establishes a one-to-one assignment between the RIS groups and the Tx-Rx pairs. Hence, the solution will be a perfect matching that has the maximum sum of edge weights over all the perfect matchings and the existence of such a perfect matching can be guaranteed by Hall's marriage theorem \cite{Hall†1987}. As a result, we have
\vspace{-2.4mm}
\begin{equation}    \label{mstar}
\mathcal{M}^{\star}=\underset{\mathcal{M}\in\mathfrak{M}}{\operatorname{\arg \max}}\sum_{e_{qk}\in\mathcal{M}} w_{qk},
\end{equation}
where $\mathfrak{M}$ denotes the set of all perfect matchings of $\mathcal{G}$ and $\mathcal{M}^{\star}$ provides the optimal solution. Hereafter, to mathematically obtain $\mathcal{M}^{\star}$ from the constructed bipartite graph $\mathcal{G}$, we use an assignment matrix $X= (x_{qk})$, whose entries are given as:
\vspace{-2.4mm}
\begin{equation}
    x_{qk}=\begin{cases}
1, & \text{if }G_q\text{ is assigned to}\: P_k,\\
0, & \text{otherwise}.
\end{cases}
\end{equation}
The one-to-one assignment constraints can be rewritten as
\vspace{-2.4mm}
\begin{equation}
    X\mathbf{1}_K=\mathbf{1}_K,
    \qquad
    X^{T}\mathbf{1}_K=\mathbf{1}_K,
    \qquad
    X\in\{0,1\}^{K\times K},
\end{equation}
where $\mathbf{1}_K\in\mathbb{R}^{K}$ denotes the vector of all ones. Hence, every feasible $X$ is a permutation matrix and corresponds to a
perfect matching of $\mathcal{G}$. By using the Frobenius inner product
$\langle A,B\rangle_F=
\sum_{q=1}^{K}\sum_{k=1}^{K}a_{qk}b_{qk},$ the ideal/practical RIS-based assignment problem can be written as
\vspace{-2.4mm}
\begin{equation}
    \max_{X}
    \left\langle W^{\mathrm{I}/\mathrm{P}},X\right\rangle_F
\end{equation}
subject to
$X\mathbf{1}_K=\mathbf{1}_K$, $X^{T}\mathbf{1}_K=\mathbf{1}_K$, and $    X\in\{0,1\}^{K\times K}.$

\section{Proposed Framework}
\subsection{Motivation for Matching Stability Analysis}

For both the ideal and the practical RIS-based models, the assignment problem have the same feasible set $\mathfrak{M}$, but with different edge-weight matrices. The optimal $\mathcal{M}^{\star}$ in \eqref{mstar} can be efficiently obtained by using the Hungarian algorithm, an existing approach that solves an assignment problem by identifying a perfect matching that maximizes the total sum of edge weights \cite{kuhn}. 
 This algorithm identifies the optimal matching for a given weight matrix, but it does not tell whether the optimal matching will change or not when the weight matrix is changed by some values. In particular, it does not explain why the optimal assignment obtained under the ideal RIS model is preserved or changes when practical reflection effects are taken into account. To address this limitation, we propose a framework that characterizes the stability of the ideal-optimal assignment under the practical RIS model.

\subsection{Assignment Stability under the Practical RIS-based Model}

As defined in \eqref{eweight}, the respective edge weight for the assignment $G_q\rightarrow P_k$ is $w_{qk}^{\rm ideal}$ and $w_{qk}^{\rm prac}$ for the ideal and practical RIS-based scenario, respectively. Moreover, for a given $k$ $(k=1,\dots,K)$, we know from \eqref{btrel} that  $0<\beta_{ki}\left( \theta_{ki} \right)\leq1$ $(i=1,\dots,N)$, i.e., we have
\vspace{-2.4mm}
\begin{equation}  \label{iprel}
    w_{qk}^{\rm ideal} \geq w_{qk}^{\rm prac} \qquad \forall \:\: q,k.
\end{equation}
Therefore, by using \eqref{eweight}, we define the practical RIS induced loss on the edge $(G_q,P_k)$ as
\vspace{-2.4mm}
\begin{equation}  \label{dqkdef}
d_{qk}=w_{qk}^{\rm ideal}-w_{qk}^{\rm prac} \geq0.
\end{equation}
This results in $W^{\rm P}
=W^{\rm I}-{D}$, where $D=(d_{qk})\in\mathbb{R_+}^{K\times K}$. Furthermore, as defined previously, \(\mathfrak{M}\) is the set of all perfect matchings between the RIS-group vertex set and the Tx-Rx pair vertex set. Hence, for $K$ RIS groups and \(K\) Tx-Rx pairs, we have $|\mathfrak{M}|=K!$. Finally, for any perfect matching \(\mathcal{M}\in\mathfrak{M}\), we define its total ideal and practical weight as
\vspace{-2.4mm}
\begin{equation}  \label{sipdef}
S_{\mathrm{I}}(\mathcal{M})
=
\sum_{e_{qk}\in\mathcal{M}}
w_{qk}^{\mathrm{ideal}} \:\: {\rm and} \:\: S_{\mathrm{P}}(\mathcal{M})
=
\sum_{e_{qk}\in\mathcal{M}}
w_{qk}^{\mathrm{prac}},
\end{equation}
respectively. Here, $S_{\mathrm{I}}(\mathcal{M})$ and $S_{\mathrm{P}}(\mathcal{M})$ denotes the sum of edge weights associated with $\mathcal{M}$ under the ideal and practical RIS-based  models, respectively. Therefore, the total practical RIS induced loss is
\vspace{-2.4mm}
\begin{equation}  \label{ldef}
L(\mathcal{M})
=
\sum_{e_{qk}\in\mathcal{M}}
d_{qk} \implies S_{\mathrm{P}}(\mathcal{M}) =
S_{\mathrm{I}}(\mathcal{M})
- L(\mathcal{M}).
\end{equation}

\subsection{Need of a Stability Theorem}

For every $\mathcal{M}\in
\mathfrak{M}\setminus
\{\mathcal{M}_{\mathrm{I}}^{\star}\}
$, where $\mathcal{M}_{\mathrm{I}}^{\star}$ is the optimal matching with respect to the ideal RIS-based scenario, we define the ``ideal matching gap'' $\Delta_{\mathrm{I}}(\mathcal{M})$ as the difference in the total edge weights between $\mathcal{M}_{\mathrm{I}}^{\star}$ and $\mathcal{M}$, i.e.,
\vspace{-2.4mm}
\begin{equation}  \label{imgdef}
    \Delta_{\mathrm{I}}(\mathcal{M})=
S_{\mathrm{I}} (\mathcal{M}_{\mathrm{I}}^{\star}) - S_{\mathrm{I}}(\mathcal{M}).
\end{equation}
Moreover, from \eqref{iprel} and \eqref{sipdef}, we obtain $S_{\rm I}(\mathcal{M}) \geq S_{\rm P}(\mathcal{M})$, which finally results in  $L(\mathcal{M}) \geq 0$ $\forall \:\: \mathcal{M}\in
\mathfrak{M}$. Note that, even in the practical RIS-based scenario,   $\mathcal{M}_{\mathrm{I}}^{\star}$ remains uniquely optimal if and only if $S_{\mathrm{P}}(\mathcal{M}_{\mathrm{I}}^{\star}) > S_{\mathrm{P}}(\mathcal{M})$ for every other competing $\mathcal{M}$. In other words, we have
\vspace{-2.4mm}
\begin{equation}  \label{newlab}
    S_{\mathrm{P}}(\mathcal{M}_{\mathrm{I}}^{\star})-S_{\mathrm{P}}(\mathcal{M}) > 0 \:\: {\rm for\:\: every} \:\: \mathcal{M}\in\mathfrak{M}\setminus
\{\mathcal{M}_{\mathrm{I}}^{\star}\}.
\end{equation}

However, the uniqueness of $\mathcal{M}_{\mathrm{I}}^{\star}$, and possibly its optimality, is lost if at least one $\mathcal{M}\in\mathfrak{M}$ exists for which \eqref{newlab} does not hold. Thus, a framework is needed to assess the validity of ideal RIS-based assignments in practical scenarios and explain any failures. This is important because considering the practical RIS-based scenario does not necessarily penalize all the feasible perfect matchings equally. Also, assignment switching occurs only when the relative loss of the ideal-optimal matching exceeds its ideal matching gap over a competitor. In this direction, the proposed theorem is as follows:


\begin{theorem}  \label{theo1}
{(Exact condition)}
Let
\vspace{-2.4mm}
\begin{equation}
    \mathcal{M}_{\mathrm{I}}^{\star}
=
\arg\max_{\mathcal{M}\in\mathfrak{M}}
S_{\mathrm{I}}(\mathcal{M})
\end{equation}
be the unique maximum-weight perfect matching under the ideal RIS-based model. By using the ideal matching gap $\Delta_{\mathrm{I}}(\mathcal{M})$ defined in \eqref{imgdef}, we state
\begin{enumerate}
\item The optimal $\mathcal{M}_{\mathrm{I}}^{\star}$ remains the unique maximum-weight perfect matching even under the practical RIS-based scenario if and only if
\vspace{-2.4mm}
\begin{equation}
L(\mathcal{M}_{\mathrm{I}}^{\star}) - L(\mathcal{M}) < \Delta_{\mathrm{I}}(\mathcal{M}),
\qquad
\forall\,
\mathcal{M}\neq\mathcal{M}_{\mathrm{I}}^{\star}.
\label{exact}
\end{equation}

\item A competing matching $\mathcal{M}$ strictly outperforms
$\mathcal{M}_{\mathrm{I}}^{\star}$ under the practical RIS-based scenario if and
only if
\vspace{-2.4mm}
\begin{equation}
L(\mathcal{M}_{\mathrm{I}}^{\star}) - L(\mathcal{M}) >
\Delta_{\mathrm{I}}(\mathcal{M}).
\end{equation}
\end{enumerate}

\end{theorem}

\begin{proof}
See Appendix A.
\end{proof}

\begin{figure}[!t]
\centering
\includegraphics[width=0.60\columnwidth]
{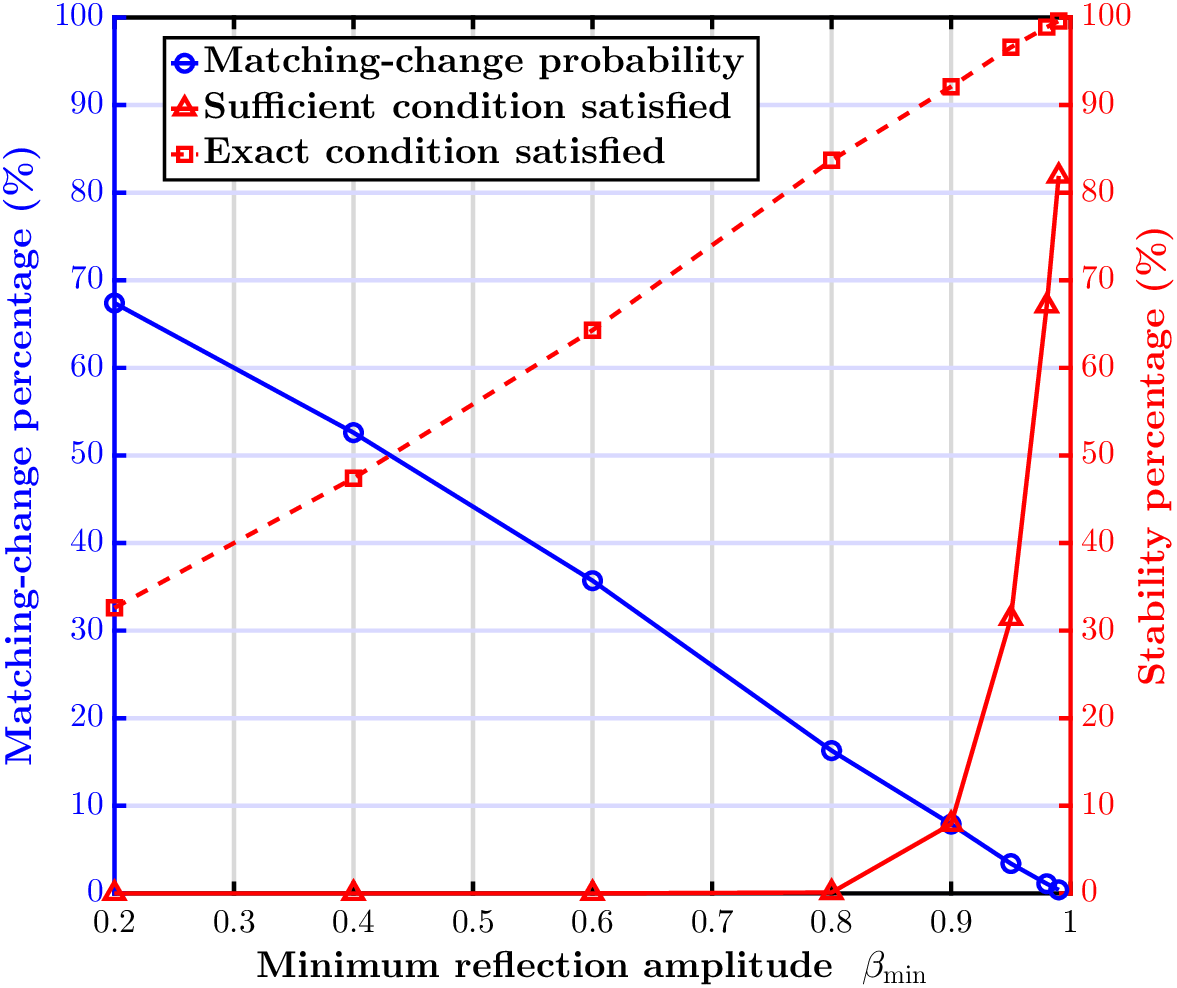}
\vspace{-2.4mm}
\caption{Impact of $\beta_{\min}$ on the matching-stability behavior.}
\label{fig:matching_change}
\vspace{-6mm}
\end{figure}

We observe that the above theorem provides much more than just the optimal assignment. While the conventional Hungarian algorithm only answers the question of \emph{which} matching is optimal for a given weight matrix, the proposed theorem addresses the complementary questions of whether the ideal RIS-based solution is robust to the practical RIS effects, why it ceases to be optimal in certain scenarios, and which competing assignments are guaranteed to outperform it under the consideration of a practical RIS-based model.

Note that sometimes, $\mathcal{M}_{\mathrm{I}}^{\star}$ may remain optimal while losing its uniqueness. The following corollary characterizes the conditions under which $\mathcal{M}_{\mathrm{I}}^{\star}$ loses uniqueness.

\begin{corollary}
If we have
\vspace{-2.4mm}
\begin{equation}
    L(\mathcal{M}_{\mathrm{I}}^{\star})-L(\mathcal{M}) \leq
    \Delta_{\mathrm{I}}(\mathcal{M}) \qquad
    \forall\,\mathcal{M}\neq\mathcal{M}_{\mathrm{I}}^{\star},
\end{equation}
and equality holds for at least one competing matching $\widehat{\mathcal{M}}$, then $\mathcal{M}_{\mathrm{I}}^{\star}$ remains optimal under the practical RIS-based scenario, but it is not unique.
\end{corollary}

\begin{proof}
  The result follows directly from Theorem 1.  
\end{proof}

The Theorem \ref{theo1} gives the exact conditions for the stability of $\mathcal{M}_{\mathrm{I}}^{\star}$. However, checking these conditions requires considering all competing perfect matchings. Hence, it is useful to have a simpler condition that guarantees stability without performing all such comparisons. The next proposition gives such a sufficient condition by using the $L(\mathcal{M}_{\mathrm{I}}^{\star})$ and min of $\Delta_{\mathrm{I}}(\mathcal{M})$.

\begin{prop} \label{cor1} {(Sufficient Condition)}
Let $\mathcal{M}_{\mathrm{I}}^{\star}$ be the unique optimal perfect matching in the ideal RIS-based scenario and we define minimum ideal matching gap
\vspace{-2.4mm}
\begin{equation}
\Delta_{\min}=
\min_{\mathcal{M}\neq\mathcal{M}_{\mathrm{I}}^{\star}}
\Delta_{\mathrm{I}}(\mathcal{M}).
\end{equation}
If $L(\mathcal{M}_{\mathrm{I}}^{\star})<\Delta_{\min}$, then $\mathcal{M}_{\mathrm{I}}^{\star}$ remains the unique optimal
perfect matching in the practical RIS-based scenario  as well.
\end{prop}

\begin{proof}
See Appendix B.
\end{proof}

\section{Numerical Results and Conclusion}

In this section, we validate our proposed framework. We consider a Rayleigh fading scenario with $4$ Tx-Rx pairs and an RIS partitioned into $4$ non-overlapping groups. Each RIS group contains $N=16$ elements, resulting in a total of $N_T=64$ RIS elements. A three-dimensional coordinate system is considered, where the Tx locations are $T_1=(0,-9, 1.5)$, $T_2=(2, -3, 1.5)$, $T_3 =(-1, 3, 1.5)$, and $T_4=(1, 9, 1.5)$ and their corresponding Rx locations are $R_1=(40, -8, 1.5)$, $R_2=(42,-2,1.5)$, $R_3=(38, 4, 1.5)$, and $R_4=(44, 9, 1.5)$. The RIS is centered at $(20,0,5)$, the Tx--Rx pairing is $P_k=(T_k,R_k)$ $\forall$ $k$, and we determine the assignment of the RIS groups to $P_1,\dots,P_4$. The path-loss at a reference distance of one meter is $L_0=40$ dB and the path-loss exponents are $\alpha_{\rm TR}=2.2$ and $\alpha_{\rm RR}=2.8$, respectively \cite{abeywickrama2020practical}. Unless otherwise specified, we consider transmission power $P_{t,k}=30$ dBm and noise power $\sigma_k^2=-90$ dBm $\forall$ $k$.

An illustrative channel realization is first used to demonstrate the optimal RIS-group assignment by considering both  the ideal and practical RIS-based reflection models. Subsequently, extensive Monte Carlo simulations verify the consistency of the proposed exact and sufficient stability conditions, i.e., Theorem $1$ and Proposition $1$, respectively. A total of $10^5$ independent Monte Carlo channel realizations are used, where for every realization, the ideal and practical edge-weight matrices are constructed as in \eqref{eweight} and the corresponding maximum-weight perfect matchings are determined. Moreover, in Fig. \ref{fig:matching_change}, the same set of channel realizations is retained while varying $\beta_{\min}$ so that the observed changes in the matching behavior are attributed solely to the practical RIS-based model rather than to different random channel samples. For the first Monte Carlo realization, the ideal and practical matchings are 
$\mathcal{M}_{\mathrm{I}}^{\star}$=$\{\{1,3\}, \{2,4\}, \{3,1\}, \{4,2\}\}$, 
and $\mathcal{M}_{\mathrm{P}}^{\star}$=$\{\{1,3\}, \{2,1\}, \{3,2\}, \{4,4\}\}$, respectively. Thus, optimal assignment changes from ideal to practical. Moreover, the sufficient condition of Proposition~\ref{cor1} is not satisfied; hence, it should be treated as inconclusive in this case.
\begin{table}[t]
\centering
\caption{Mean stability metrics for different values of $\beta_{\min}$.}
\vspace{-3.6mm}
\label{table:stability_beta}
\renewcommand{\arraystretch}{1}
\setlength{\tabcolsep}{7.2pt}
\begin{tabular}{ccc}
\toprule
$\beta_{\min}$ &
Mean $\Delta_{\min}$ &
Mean $L(\mathcal{M}_{I}^{\star})$ \\
\midrule
0.20 & 0.0909 & 1.5202 \\
0.40 & 0.0909 & 1.1947 \\
0.60 & 0.0909 & 0.8261 \\
0.80 & 0.0909 & 0.4248 \\
0.90 & 0.0909 & 0.2148 \\
0.95 & 0.0909 & 0.1080 \\
0.98 & 0.0909 & 0.0433 \\
0.99 & 0.0909 & 0.0217 \\
\bottomrule
\end{tabular}
\vspace{-6mm}
\end{table}

Fig. \ref{fig:matching_change} validates the proposed exact matching-stability condition in Theorem~\ref{theo1} and its sufficient-condition in Proposition~\ref{cor1}. For every $\beta_{\min}$, the exact stability condition characterizes whether $\mathcal{M}_I^{\star}$ remains the unique optimum under the practical RIS model. The number of realizations satisfying the condition-1 of Theorem 1 coincides exactly with the number of realizations for which $\mathcal{M}_I^{\star}$ remains uniquely optimal, with zero theorem violations over all the tested cases. In addition, Table \ref{table:stability_beta} shows that the mean loss $L(\mathcal{M}_I^{\star})$ decreases monotonically with increasing $\beta_{\min}$ ($\because$ the practical RIS approaches its ideal counterpart as $\beta_{\min} \rightarrow 1$) whereas $\Delta_{\min}$ remains constant as it is independent of $\beta_{\min}$. The figure also depicts that the probability of the ideal-optimal matching remaining unchanged increases from $32.6\%$ to $99.6\%$, while the matching-change probability decreases from $67.4\%$ to only $0.4\%$.

Note that, Proposition~\ref{cor1} based on the sufficient condition $L(\mathcal{M}_I^{\star})<\Delta_{\min}$ is conservative. However, it provides a direct comparison rule to check the retention of the ideal-optimal solution without comparing it with every element of the feasible set. The percentage of realizations satisfying the Proposition~\ref{cor1} increases from $0\%$ to $81.9\%$ as $\beta_{\min}$ increases  
with zero sufficient-condition violations, as shown by the right vertical axis of Fig \ref{fig:matching_change}. These observations highlight the complementary nature of the results: Theorem~\ref{theo1} provides an exact stability characterization, whereas Proposition~\ref{cor1} gives a simpler sufficient condition for retaining the ideal RIS-based optimal assignment even in the practical RIS-based scenario. 

\begin{figure}[!t] 
\centering
\includegraphics[width=0.6\columnwidth]
{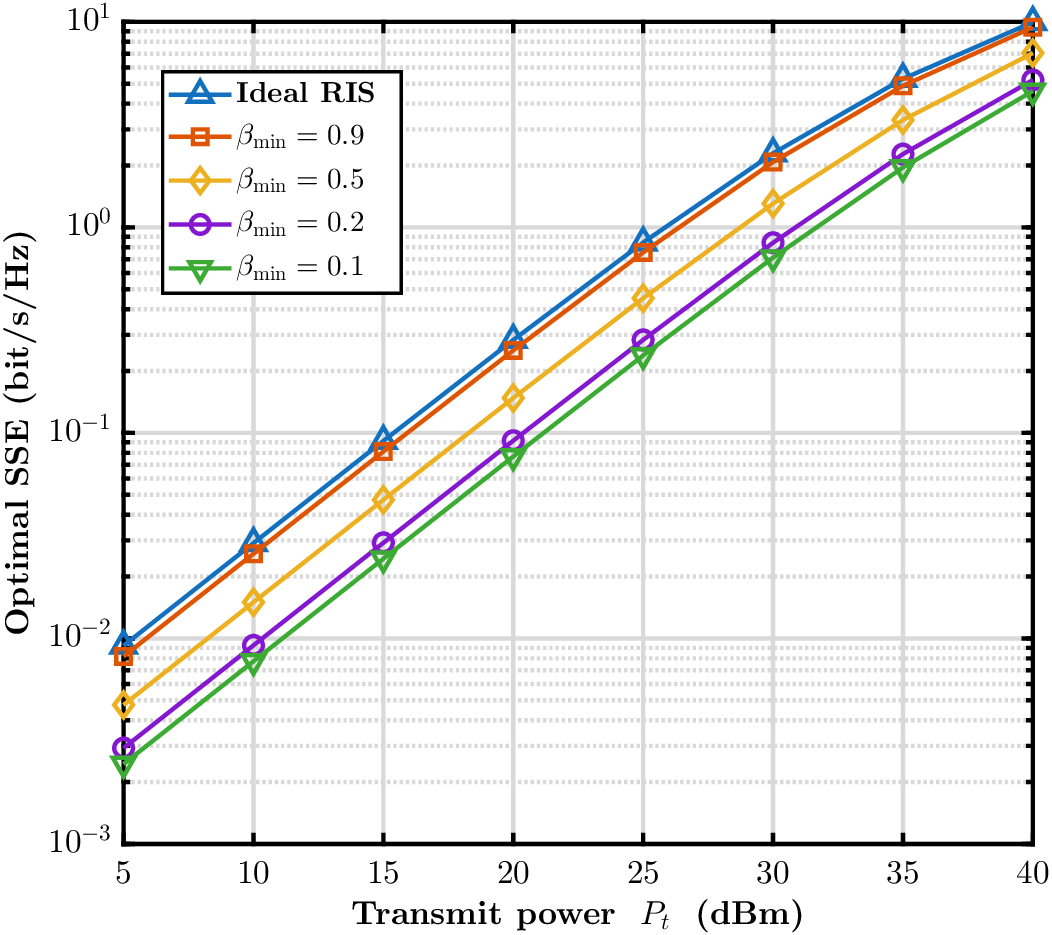}
\vspace{-2.4mm}
\caption{Optimal SSE versus transmission power.}
\label{transP}
\vspace{-6mm}
\end{figure}

Fig.~ \ref{transP} demonstrates the impact of the RIS model on the system performance in terms of the optimal SSE. Specifically, with various transmission power, we evaluate  both $S_I\left(\mathcal{M}_{I}^{\star}\right)$ and $S_P\left(\mathcal{M}_{P}^{\star}\right)$ for different $\beta_{\min}$ values. 
We observe that, irrespective of the considered RIS model, the optimal SSE increases with the transmission power. This is because the received SNR increases proportionately with the transmission power, which in turn, enhances the SSE. A performance gap is observed between the ideal and practical RIS-based scenarios, which is particularly noticeable at $\beta_{\min}=0.1$ between $S_I(\mathcal{M}_{I}^{\star})$ and $S_P(\mathcal{M}_{P}^{\star})$, and it decreases as $\beta_{\min}\rightarrow1$. This is because $\beta_{\min} \rightarrow 1$ implies a transition from the practical to the ideal RIS-based scenario.

Therefore, we proposed a graph theoretic framework for the grouping-based practical RIS-aided multi-user networks. We investigated the user-RIS association problem, which results in maximizing the system performance. We also demonstrated that the ideal RIS-based model significantly overestimates the system performance as compared to its practical counterpart.

\section*{Appendix}

\vspace{-2mm}

\subsection{Proof of Theorem 1} \label{app1}

\vspace{-1mm}
By using \eqref{sipdef} and \eqref{ldef}, we obtain $S_{\mathrm{P}}(\mathcal{M}) = \sum_{e_{qk}\in\mathcal{M}}
w_{qk}^{\mathrm{prac}}=S_{\mathrm{I}}(\mathcal{M})
- L(\mathcal{M})$
for any perfect matching $\mathcal{M}\in\mathfrak{M}$. Accordingly, we consider an arbitrary competing matching $\mathcal{M}\neq\mathcal{M}_{\mathrm{I}}^{\star}$ to obtain
\vspace{-2.4mm}
\begin{align}  \label{lsp}
& L(\mathcal{M}_{\mathrm{I}}^{\star}) - L(\mathcal{M}) \nonumber \\
&= \left( S_{\mathrm{I}}\left( \mathcal{M}_{\mathrm{I}}^{\star} \right) -  S_{\mathrm{P}}\left( \mathcal{M}_{\mathrm{I}}^{\star} \right) \right) - \left( S_{\mathrm{I}}\left( \mathcal{M} \right) -  S_{\mathrm{P}}\left( \mathcal{M} \right) \right) \nonumber \\
&=\Delta_{\mathrm{I}}(\mathcal{M}) - \left( S_{\mathrm{P}}(\mathcal{M}_{\mathrm{I}}^{\star}) - S_{\mathrm{P}}(\mathcal{M}) \right).
\end{align}
Note that, $\mathcal{M}_{\mathrm{I}}^{\star}$ remains strictly better than $\mathcal{M}$ even in the practical RIS-based scenario if and only if we have
\vspace{-2.4mm}
\begin{align}
\!\!\!\!S_{\mathrm{P}}(\mathcal{M}_{\mathrm{I}}^{\star}) > S_{\mathrm{P}}(\mathcal{M}) & \implies S_{\mathrm{P}}(\mathcal{M}_{\mathrm{I}}^{\star}) - S_{\mathrm{P}} (\mathcal{M}) >0 \nonumber \\
& \implies L(\mathcal{M}_{\mathrm{I}}^{\star}) - L(\mathcal{M}) < \Delta_{\mathrm{I}}(\mathcal{M}).
\end{align}
This completes the first part. Next, a competing matching $\mathcal{M}$ strictly outperforms $\mathcal{M}_{\mathrm{I}}^{\star}$ under the practical RIS-based model if and only if $S_{\mathrm{P}}(\mathcal{M}_{\mathrm{I}}^{\star}) < S_{\mathrm{P}}(\mathcal{M})$, i.e.,
\vspace{-2.4mm}
\begin{align}
 \!\! \!\! S_{\mathrm{P}}(\mathcal{M}_{\mathrm{I}}^{\star}) \! - \! S_{\mathrm{P}} (\mathcal{M}) \! < \! 0 \implies \!\! L(\mathcal{M}_{\mathrm{I}}^{\star}) \! - \! L(\mathcal{M}) > \Delta_{\mathrm{I}}(\mathcal{M}),
\end{align}
which proves the second statement.

\vspace{-3.2mm}

\subsection{Proof of Proposition 1}

\vspace{-1mm}

Based on \eqref{dqkdef} and \eqref{ldef}, we have $L(\mathcal{M})\geq 0$, i.e.,
\vspace{-2.4mm}
\begin{equation}  \label{eq1}
L(\mathcal{M}_{\mathrm{I}}^{\star})-L(\mathcal{M})\leq
L(\mathcal{M}_{\mathrm{I}}^{\star}).
\end{equation}
Moreover, by the definition of $\Delta_{\min}$, we get
\vspace{-2.4mm}
\begin{equation}  \label{eq2}
    \Delta_{\min}
\leq
\Delta_{\mathrm{I}}(\mathcal{M})
\qquad
\forall\,
\mathcal{M}\neq\mathcal{M}_{\mathrm{I}}^{\star}.
\end{equation}
Thus, if $L(\mathcal{M}_{\mathrm{I}}^{\star})<\Delta_{\min}$, by using \eqref{eq1} and \eqref{eq2}, we obtain
\vspace{-2.4mm}
\begin{equation}
    L(\mathcal{M}_{\mathrm{I}}^{\star})
-
L(\mathcal{M})
\leq
L(\mathcal{M}_{\mathrm{I}}^{\star})
<
\Delta_{\min}
\leq
\Delta_{\mathrm{I}}(\mathcal{M})
\end{equation}
for every competing $\mathcal{M}$. This, in turn, results in $L(\mathcal{M}_{\mathrm{I}}^{\star})
-
L(\mathcal{M})
<
\Delta_{\mathrm{I}}(\mathcal{M})$ $\forall$ $
\mathcal{M}\neq\mathcal{M}_{\mathrm{I}}^{\star},$
which is precisely the condition 1 of Theorem 1. Therefore, if $L(\mathcal{M}_{\mathrm{I}}^{\star})<\Delta_{\min}$ holds, $\mathcal{M}_{\mathrm{I}}^{\star}$ remains the unique optimal perfect matching even in the practical RIS-based scenario.

\vspace{-2.4mm}
\bibliographystyle{IEEEtran}
\bibliography{references}

\begin{thebibliography}{10}
\providecommand{\url}[1]{#1}
\csname url@samestyle\endcsname
\providecommand{\newblock}{\relax}
\providecommand{\bibinfo}[2]{#2}
\providecommand{\BIBentrySTDinterwordspacing}{\spaceskip=0pt\relax}
\providecommand{\BIBentryALTinterwordstretchfactor}{4}
\providecommand{\BIBentryALTinterwordspacing}{\spaceskip=\fontdimen2\font plus
\BIBentryALTinterwordstretchfactor\fontdimen3\font minus
  \fontdimen4\font\relax}
\providecommand{\BIBforeignlanguage}[2]{{%
\expandafter\ifx\csname l@#1\endcsname\relax
\typeout{** WARNING: IEEEtran.bst: No hyphenation pattern has been}%
\typeout{** loaded for the language `#1'. Using the pattern for}%
\typeout{** the default language instead.}%
\else
\language=\csname l@#1\endcsname
\fi
#2}}
\providecommand{\BIBdecl}{\relax}
\BIBdecl

\bibitem{8811733}
Q.~Wu and R.~Zhang, ``Intelligent reflecting surface enhanced wireless network
  via joint active and passive beamforming,'' \emph{IEEE Trans. Wireless
  Commun.}, vol.~18, no.~11, pp. 5394--5409, Nov. 2019.

\bibitem{9140329}
M.~Di~Renzo \emph{et~al.}, ``Smart radio environments empowered by
  reconfigurable intelligent surfaces: How it works, state of research, and the
  road ahead,'' \emph{IEEE J. Sel. Areas Commun.}, vol.~38, no.~11, pp.
  2450--2525, Nov. 2020.

\bibitem{tnse}
L.~Sau, P.~Mukherjee, and S.~C. Ghosh, ``Priority-aware grouping-based multihop
  routing scheme for {RIS}-assisted wireless networks,'' \emph{IEEE Trans.
  Network Sci. Eng.}, vol.~12, no.~2, pp. 1172--1185, Mar. 2025.

\bibitem{partition3}
Y.~Yang, B.~Zheng, S.~Zhang, and R.~Zhang, ``Intelligent reflecting surface
  meets {OFDM}: Protocol design and rate maximization,'' \emph{IEEE Trans.
  Commun.}, vol.~68, no.~7, pp. 4522--4535, Jul. 2020.

\bibitem{chan}
K.-L. Chan \emph{et~al.}, ``Beamforming and load-balanced user association in
  {RIS}-aided {mmWave} systems via adaptive attention graph neural networks,''
  \emph{IEEE Trans. Wireless Commun.}, vol.~25, pp. 5781--5796, 2026.

\bibitem{sautcom}
L.~Sau, P.~Mukherjee, and S.~C. Ghosh, ``Generalized group selection strategies
  for self-sustainable {RIS}-aided communication,'' \emph{IEEE Trans. Commun.},
  vol.~74, pp. 6584--6598, 2026.

\bibitem{abeywickrama2020practical}
S.~Abeywickrama, R.~Zhang, Q.~Wu, and C.~Yuen, ``Intelligent reflecting
  surface: Practical phase shift model and beamforming optimization,''
  \emph{IEEE Trans. Commun.}, vol.~68, no.~9, pp. 5849--5863, Sep. 2020.

\bibitem{wcl3}
P.~Mukherjee \emph{et~al.}, ``Chaotic noncoherent {SWIPT} in multi-functional
  {RIS}-aided systems,'' \emph{IEEE Wireless Commun. Lett.}, vol.~15, pp.
  6--10, 2026.

\bibitem{bipartite}
D.~B. West, \emph{Introduction to Graph Theory}.\hskip 1em plus 0.5em minus
  0.4em\relax Pearson Education, Inc., 2001.

\bibitem{Hall†1987}
P.~Hall, ``On representatives of subsets,'' \emph{J. London Math. Soc.}, vol.
  s1-10, no.~1, pp. 26--30, Jan. 1935.

\bibitem{kuhn}
H.~W. Kuhn, ``The {H}ungarian method for the assignment problem,'' \emph{Nav.
  Res. Logist. Q.}, vol.~2, no. 1-2, pp. 83--97, Mar. 1955.

\end{thebibliography}

\end{document}